\documentclass[12pt]{iopart}

\usepackage{iopams}  
\usepackage{graphicx}
\pdfoutput=1

\begin{document}

\title[]{Moulding the flow of surface plasmons using conformal and quasiconformal mapping.}
\author{P A Huidobro$^1$, M L Nesterov$^1$ \footnote{A. Ya. Usikov Institute for Radiophysics and
Electronics, NAS of Ukraine, 12 Academician Proskura Street, 61085
Kharkov, Ukraine}, L Mart\'in-Moreno$^2$, F J Garc\'ia-Vidal$^1$}
\address{$^1$ Departamento de F\'isica Te\'orica de la Materia
Condensada, Universidad Aut\'onoma de Madrid, E-28049 Madrid,
Spain}
\address{$^2$ Instituto de Ciencia de
Materiales de Arag\'on and Departamento de F\'isica de la Materia Condensada,
CSIC--Universidad de Zaragoza, E-50009 Zaragoza, Spain}
\ead{paloma.arroyo@uam.es}

\begin{abstract}
In this paper we analyze how Transformation Optics recipes can be applied to control the flow of surface plasmons on metal-dielectric interfaces. We study in detail five different examples: a cylindrical cloak, a beam shifter, a right-angle bend, a lens and a ground-plane cloak.  First, we demonstrate that only the modification of the electric permittivity and magnetic permeability in the dielectric side can lead to almost perfect functionalities for surface plasmons.  We also show that, thanks to the quasi two-dimensional character of surface plasmons and its inherent polarization, the application of conformal and quasiconformal mapping techniques allows the design of plasmonic devices in which only the isotropic refractive index of the dielectric film needs to be engineered.  
\end{abstract}

\vspace{2pc}
\noindent{\it Keywords}: Transformation Optics, Surface Plasmon Polaritons, optical conformal mapping, quasiconformal mapping.

\maketitle

\section{Introduction}
Transformation Optics (TO) \cite{PendrySci06,LeonhardtPhilbin06} has attracted great interest in the past few years due to its potential ability to control the flow of light. It establishes a procedure that provides us with expressions for the electromagnetic (EM) material parameters, dielectric permittivity and magnetic permeability, that need to be implemented in order to obtain a medium where EM waves propagate along desired trajectories. It is based on the mathematical equivalence between coordinate transformations and EM material parameters. As Maxwell's equations are form-invariant under coordinate transformations, a distorted geometry in an empty space can be interpreted as a medium in a Cartesian flat space \cite{Pendry96}. A wide variety of optical devices has been designed by means of TO, such as invisibility cloaks \cite{PendrySci06, Leonhardt06}, beam shifters and dividers \cite{Rahm08}, beam bends \cite{Smith08,Kwon}, waveguide bends \cite{RahmRoberts}, field concentrators \cite{RahmSchurig08}, imaging devices \cite{Schurig07,Kildishev} or polarization splitters and rotators \cite{{Kwon08}}. The permittivity and permeability tensors required by the TO technique are usually implemented using metamaterials, artificial materials that allow to effectively mimic EM responses by repeating structured unit cells at a subwavelength scale \cite{Ziolkowski,Ramakrishna}. However, these parameters are generally anisotropic and inhomogeneous, so metamaterials usually rely on resonating structures to tailor EM parameters, a feature that limits the wavelength range of operation and causes losses. Conformal \cite{Leonhardt06} and quasiconformal \cite{JensenLi08} coordinate transformations have been proposed to overcome this problem, since they reduce anisotropy to a minimum yielding the possibility of fabricating transformation media out of isotropic dielectrics. Consequently, devices based on conformal transformations can be inherently broadband at optical frequencies. While conformal mappings are usually obtained analytically \cite{Schmiele}, quasiconformal mappings rely on numerical grid generation techniques. Recently, a method based on solving Laplace's equation with the appropriate boundary conditions has been proposed \cite{Chang,Hu} to generate quasiconformal mappings for different optical elements such as cloaks or arbitrary waveguides. 
 
It has been recently shown that TO recipes can be applied to efficiently mould the flow of surface plasmon polaritons (SPPs) \cite{Huidobro, Liu10,Kadic,Renger,Zentgraf}. SPPs are highly confined waves that propagate along a metal-dielectric interface. They are evanescent in the direction perpendicular to the interface and their field decays exponentially in the metal and dielectric sides. SPPs arise from the resonant interaction of light with the free electrons of the metal and result in a collective oscillation \cite{Barnes03,Maier05}. SPPs are considered as good candidates for photonic circuitry as they provide large field enhancements at subwavelength scales \cite{Bozhevolnyi08,Gramotnev10}. On the other hand, within the field of Metamaterials, plasmonic structures have already been developed to achieve cloaking. They include, for instance, the cloaking from an SPP of an object surrounded by a circular array of point scatterers \cite{Maradudin}, the experimental verification of a plasmonic shell capable of cloaking subwalength objects \cite{Engheta09} or a two-dimensional cloak for SPPs consisting of layers supporting plasmonic modes with group velocity of opposite signs \cite{Smolyaninov}. However, it is interesting to note that using the TO tools to manipulate SPPs is a very general technique that can yield different optical devices and new functionalities. Furthermore, as it has been shown in previous works \cite{Huidobro, Liu10}, only the dielectric material next to the metal needs to be structured, while modifying the metal properties is not required in most cases. 

In this manuscript we consider the main features of TO as applied to SPPs and exemplify the theory by means of different TO devices for SPPs. We verify the results using the Finite Element Method solver COMSOL MULTIPHYSICS. First, we introduce the general methodology of TO in \sref{sect:TOforSPPs} and apply it to the case of SPPs. In particular, we study the functionality of four different potential plasmonic devices (a cylindrical cloak, a parallel shifter, a right-angle bend and a lens) at a range of wavelengths through the optical range of the spectrum. Second, we consider realistic models based on conformal and quasiconformal mappings in \sref{sect:conformalAndQuasiconf}. We show that for these cases only isotropic refractive indexes need to be implemented and we apply it to the parallel shifter, the right-angle bend, the lens and a ground-plane cloak for SPPs. Finally, \sref{sect:conclusions} summarizes the main contributions of our work.    

%

\section{Transformation Optics for Plasmonics} \label{sect:TOforSPPs}
The TO framework provides tools for the design of EM media with desired functionalities. It is based on the fact that Maxwell's equations are left invariant if a mapping is performed from an empty space with a distorted geometry (virtual space) to a medium characterized by $\hat{\epsilon}$ and $\hat{\mu}$ placed in a flat space (physical space). A two-dimensional (2D) transformation $x^i = x^i(x^{i'})$ between the virtual system, $x^{i'}=\{x',y'\}$, and the physical system, $x^i=\{x,y\}$, is described by means of its Jacobian matrix:
\begin{equation}
	\widehat{A}= \left(\frac{\partial x^i}{\partial x^{i'}}\right)
\end{equation}
where the index $i=1,2$. As Maxwell's equations are form-invariant, the empty-space equations in physical space can be interpreted as macroscopic equations for a medium placed in virtual space and characterized by the following EM parameters:
\begin{equation} \label{eq:parameters}
	\hat{\epsilon}_{TO}=\epsilon_0 \frac{\widehat{A}\widehat{A}^T}{\det \widehat{A}} \mbox{;} \ \ \ \hat{\mu}_{TO}=\mu_0 \frac{\widehat{A}\widehat{A}^T}{\det \widehat{A}}
\end{equation}
where $\epsilon_0$ and $\mu_0$ are the background EM parameters of the physical space. On the other hand, we can relate the metric tensor $\widehat{G}= \left(g_{ij}\right)$ of the coordinate transformation to the inverse Jacobian matrix:
\begin{equation} \label{eq:G}
	\widehat{G}=\frac{\partial x^{k}}{\partial x^i}\frac{\partial x^{k}}{\partial x^j}=\left(\widehat{A}^{-1}\right)^T\widehat{A}^{-1}
\end{equation}
This allows us to re-express $\hat{\epsilon}_{TO}$ and $\hat{\mu}_{TO}$ in terms of the inverse metric tensor, $\widehat{G}^{-1}=\widehat{A}\widehat{A}^T$, of the transformation:
\begin{equation} \label{eq:parametersG}
	\hat{\epsilon}_{TO}=\epsilon_0\sqrt{g} \widehat{G}^{-1} \mbox{;} \ \ \ 	\hat{\mu}_{TO}=\mu_0\sqrt{g} \widehat{G}^{-1} 
\end{equation}
where $g$ is the determinant of the metric, $g=1 / (\det{\widehat{A}})^2$. In general, the metric tensor of any 2D coordinate transformation is a $2\times2$ symmetric matrix. Then, as it can be readily seen from \eref{eq:parametersG}, $\hat{\epsilon}_{TO}$ and $\hat{\mu}_{TO}$ are generally anisotropic. Moreover, they are inhomogeneous and vary throughout space according to the coordinate transformation. 

Since TO relies on Maxwell's equations, it applies to all kinds of EM waves and in particular to SPPs. However, we need to bear in mind the fact that SPPs are surface waves propagating along a metal-dielectric interface with a field distribution that is evanescent in the direction perpendicular to the interface. Therefore, an infinite number of 2D coordinate transformations should be performed, in principle, in planes parallel to the metal-dielectric interface in order to operate over the whole plasmonic field. As  the SPP field extends into the dielectric and the metal, these 2D coordinate transformations should be implemented at both sides of the interface. Due to the fact that the SPP decay length in the metal (skin depth) is of a few tens of nanometers at optical wavelengths, a manipulation of the metal properties would be needed at a nanometer scale. However, it has been shown \cite{Huidobro} that a more feasible approach that consists of of modifying the EM properties of the dielectric side only  indeed gives very accurate results, while control over the metal is not actually required. To illustrate the direct application of TO to SPPs and the impact of manipulating only the dielectric side, we have studied the dependence of the efficiency on the wavelength for four different TO devices for SPPs: a cylindrical cloak, a parallel shifter, a right-angle bend and a lens. 

\begin{figure}[ht] \centering
\includegraphics{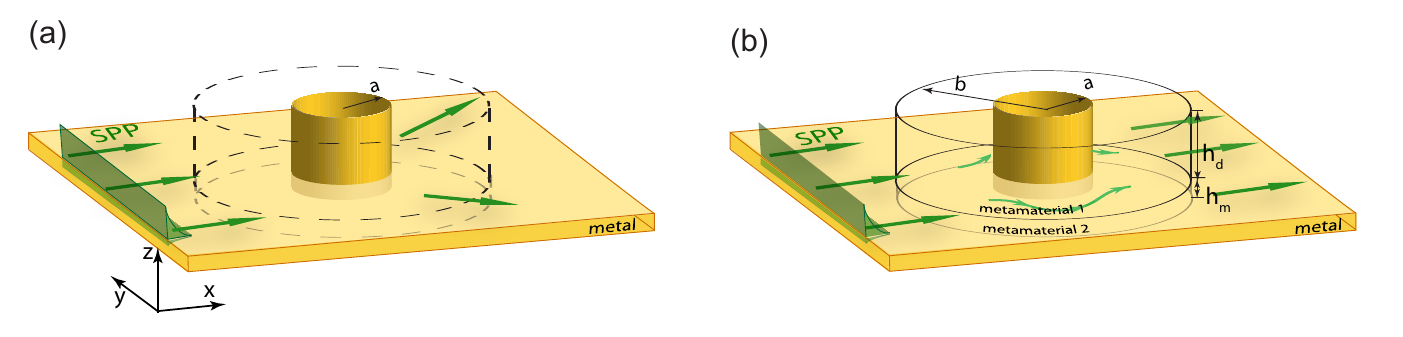} 
	\caption{Transformation Optics is used to cloak a cylinder to an SPP. (a) An SPP is propagating along a gold-vacuum interface and scatters on a metallic cylinder of radius $a$. The plasmonic field intensity, shown schematically, decays exponentially away from the interface. (b) Scattering losses are suppressed by placing a cloak of outer radius $b$ around the cylinder. The optical parameters needed to cloak the cylinder are calculated using TO and implemented in a cylindrical shell $a<r<b$ of height $h_d$ in the dielectric side ({\itshape metamaterial 1}) and in a shell of height $h_m$ in the metal ({\itshape metamaterial 2}). As the decay length of the SPP is larger in the dielectric than in the metal and most part of the SPP energy is carried in the dielectric, $h_m$ is much smaller than $h_d$.}
    \label{fig:3Dcloak}
\end{figure}

\subsection{Cylindrical cloak}
\begin{figure}[ht] \centering
\includegraphics{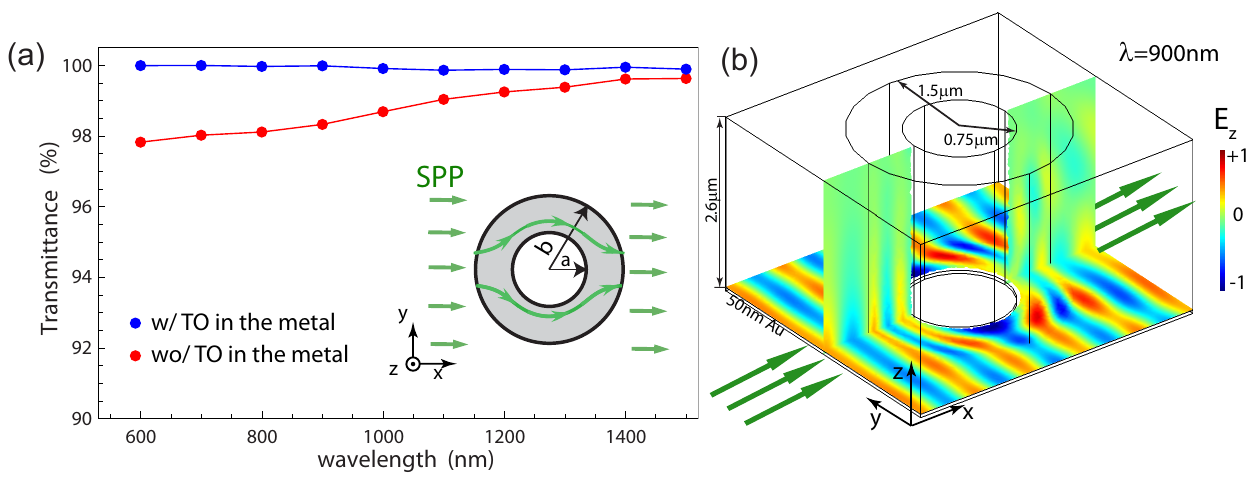} 
	\caption{Simulation results of a 3D cylindrical cloak for an SPP. (a) Transmittance (ratio of the incident power flow to the transmitted power flow) through the cloak for SPPs at different wavelengths. The blue and red lines show the case with and without placing a transformation medium in the metal side respectively. The in-plane size of the cloak is scaled with the wavelength $a=b/2=5/6\lambda$. The height of the cloak in the dielectric side, $h_d$, is much larger than the vacuum decay length at a given wavelength and is chosen for convergence. The height of the cloak in the metal side, $h_m$, is $50$ nm high at all wavelengths since the skin depth of the SPP is very similar over the wavelength range and amounts to $20-30$ nm. The inset panel shows the geometry layout from a top view. (b) z-component of the electric field of an SPP at $900$ nm travelling through a cloak without applying TO in the metal side.}
    \label{fig:CylCloak}
\end{figure}

As a first example of TO applied to plasmonics, let us consider a three-dimensional (3D) cylindrical cloak for SPPs travelling along the interface between a metal characterized by permittivity $\epsilon_m$ and vacuum \cite{Huidobro}. \Fref{fig:3Dcloak}(a) shows an SPP that scatters with a metallic cylinder of radius $a$ placed on the surface. A cloak placed around the cylinder, see \fref{fig:3Dcloak}(b), renders the object invisible by suppressing any scattering from it. A 2D invisibility cloak was the first proposal of TO applications \cite{PendrySci06, Leonhardt06} as well as the first experimental realization \cite{Schuring06}. The cloak is designed by taking a physical space that has a hole of radius $a$ surrounded by a compressed region of radius $b$ and then making a transformation to a Cartesian undistorted space. A wave travelling in physical space will follow trajectories that surround the hole so that it appears to have travelled in straight lines as it would in the virtual - undistorted - space. In order to design a 3D invisibility cloak for SPPs we take a plane parallel to the metal surface and make the 2D radial transformation $r(r')=b/(b-a)r'+a$ in the shell $a<r<b$ (see the geometry of the cloak in the inset of \fref{fig:CylCloak}(a)). In Cartesian coordinates, the transformation reads: 
\begin{equation} \label{eq:CylCTransf}
x=r(r')\frac{x'}{r'}\ \mbox{;} \ \ y=r(r')\frac{y'}{r'} \ \mbox{;} \  \ z=z'
\end{equation}
According to the TO equation \eref{eq:parametersG}, the EM parameters of the transformation medium are the following:
\begin{equation}
\label{eq:CylCparameters}
\hat{\epsilon}_{TO}/\epsilon_0=\hat{\mu}_{TO}/\mu_0= \left( \begin{array}{ccc}
   \epsilon_r \frac{x^2}{r^2}+\epsilon_\theta\frac{y^2}{r^2}&
   ( \epsilon_r-\epsilon_\theta)\frac{xy}{r^2} & 0\\
   ( \epsilon_r-\epsilon_\theta)\frac{xy}{r^2}& 
   \epsilon_r \frac{y^2}{r^2}+\epsilon_\theta\frac{x^2}{r^2}&0\\
   0&0&\epsilon_z
\end{array}\right)
\end{equation}
where $\epsilon_r=\mu_r=(r-a)/r$, $\epsilon_\theta=\mu_\theta=r/(r-a)$ and $\epsilon_z=\mu_z=[b/(b-a)]^2(r-a)/r$. These parameters are highly anisotropic and approach singular values at the inner radius of the cloak ($\epsilon_{xx}\mbox{,}\ \epsilon_{yy} \to +\infty$, $\epsilon_{xy}\to-\infty$, $\epsilon_{z}\to0$). If we want to operate on an SPP, we need to cloak the field in the dielectric and metal sides as illustrated in \fref{fig:3Dcloak}(b). In the dielectric side, we implement the parameters, $\hat{\epsilon}=\hat{\epsilon}_{TO}$ and $\hat{\mu}=\hat{\mu}_{TO}$, in a cylindrical shell of radii $a$ and $b$ with a height, $h_d$, larger than the vacuum decay length of the SPP. On the other hand, in the metal side the EM tensors need to be modified, $\hat{\epsilon}=\hat{\epsilon}_{TO}\cdot\epsilon_m$ and $\hat{\mu}=\hat{\mu}_{TO}$, in a cylindrical shell of the same radii but with a height $h_m$ larger than the skin depth. If we cloaked its entire field, an incident SPP would emerge at the other side as if it had travelled through free space.

\Fref{fig:CylCloak} shows the results obtained from 3D simulations. To study the effectiveness of the plasmonic device we have considered the transmittance of SPPs through the cloak and its scattering properties. We have measured the transmittance from one side of the cloak to the other as the ratio of the incident to the transmitted power flow. In order to isolate the effect of the TO medium, losses have been neglected throughout this work by setting the imaginary part of the metal dielectric constant to zero.  Since losses are absent, $100\%$ transmittance means that the whole plasmonic field is guided around the cylinder without any back-scattering. In the simulations, an SPP at a wavelength between $600$ and $1500$ nm and travelling through a gold-vacuum interface, impinges onto a cloak of size $a=b/2=5/6\lambda$. The transmittance spectrum is plotted in \fref{fig:CylCloak}(a) for two cases: with and without introducing a metamaterial in the metal side (blue and red lines respectively). The transmittance of the cloak in the ideal situation (blue line) where TO parameters are implemented in the metal along with the dielectric material is higher than $99\%$ over the whole wavelength range. On the other hand, the transmittance when the metal is leaved untouched (red line) increases from $97\%$ at $600$ nm to $99\%$ at $1500$ nm. As the wavelength increases, the blue and red curves lie closer. The reason for this is that while the skin depth is approximately constant over the whole spectrum (around $20$-$30$ nm), the vacuum decay length increases with the wavelength, varying from $250$ nm for an SPP at $600$ nm to $4$ $\mu$m at $1500$ nm. Thus, for shorter wavelengths, the part of the SPP energy carried within the metal is larger. However, although the effectiveness of the cloak is affected by the lack of cloaking the field that resides in the metal, it is very good even at short wavelengths. The performance of the cloak for an SPP at $\lambda=900$ nm is illustrated in \fref{fig:CylCloak}(b), which shows how the SPP field is smoothly guided around the metallic cylinder.

Finally, a study of the scattering properties is needed for a complete analysis of the invisibility cloak. By inspection of the scattered field distribution we have seen that, when the TO recipe is not implemented in the metallic side, around $20\%$ of the field amplitude is scattered in the forward direction. As a result, the wavefront of the SPP when passing through the simplified cloak is slightly modified. In accordance to \cite{Qiu}, phase distortions occur for the simplified cloak and perfect cloaking is not achieved. Nevertheless, as seen in \fref{fig:CylCloak}, most of the plasmonic field is guided around the cylinder. On the other hand, we have checked the scattering properties of all the devices shown in this work. We have found that these effects are only relevant in the case of the cylindrical cloak. For the sake of brevity, we mention this fact here and not when discussing each device. Therefore, in the rest of the paper, we present the transmittance and the images of the EM field as criteria for quantifying the performance of the SPP-devices.

\subsection{SPP parallel shifter}
\begin{figure}[!hb] \centering
\includegraphics{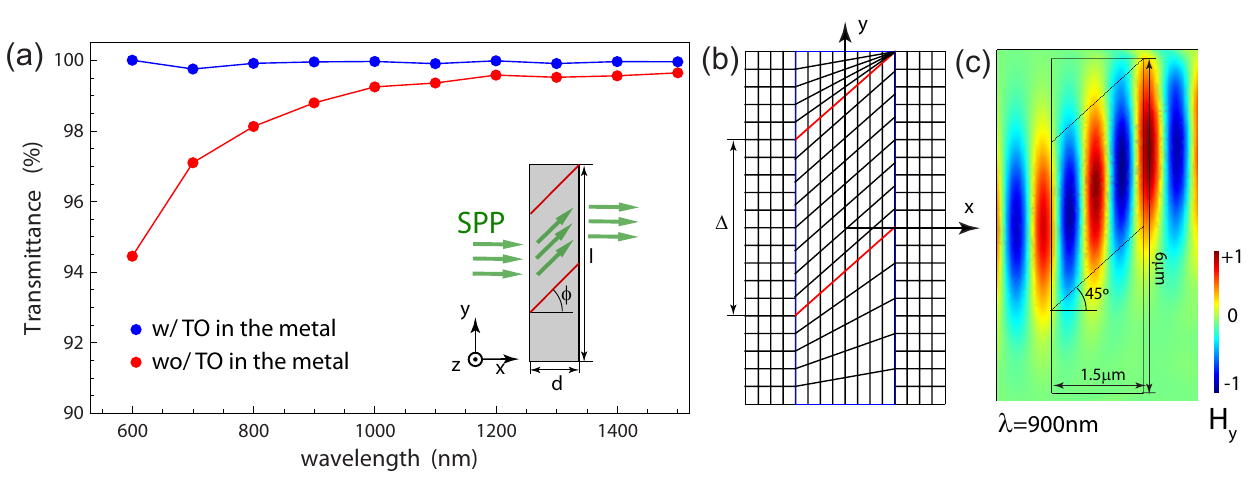} 
	\caption{3D simulation results of a parallel shifter. A gaussian SPP of width $\Delta$ impinges on the device and experiences a shift of $\phi=45^\circ$. (a) Transmittance curve of the shifter for SPPs at different wavelengths. The blue and red lines show the case with and without placing a transformation medium in the metal side, respectively. The TO parameters are implemented in a slab sized $d\times l \times h_d$ in the dielectric side and, only for the blue line, in a slab $d\times l \times h_m$ within the metal (see the inset panel for geometry parameters). The transmittance through the shift is measured as the ratio of the power flow that impinges the entrance face within $-\Delta/2<y<\Delta/2$ to the power flow that exits within $0<y<\Delta$. (b) Coordinate transformation given by \eref{eq:ShifterTransf}. (c) Top view of the magnetic field component of an SPP at $900$ nm travelling through a shifter without applying TO in the metal side. Note that the two lines at $45^\circ$ are simply a guide to the eye and do not represent a physical boundary.}
    \label{fig:Shifter}
\end{figure}

Another potential plasmonic device is the parallel shifter \cite{Huidobro}, which acts on an incident beam by translating it perpendicularly to its propagation direction. To design a 2D shifter, we take a rectangular region sized $d\times l$ (see inset panel of \fref{fig:Shifter}(a)) and map a Cartesian grid to a grid that is tilted  at an angle $\phi$ \cite{Rahm08}:
\begin{equation} \label{eq:ShifterTransf}
	x=x'\ \mbox{;} \ \ y=y'+a(x'+d) \ \mbox{;} \  \ z=z'
\end{equation}
where $a$ determines the tilt angle $a=\tan\phi$. The map of the transformation is illustrated in \fref{fig:Shifter}(b). In contrast to the transformation leading to the cylindrical cloak \eref{eq:CylCTransf}, the transformation \eref{eq:ShifterTransf} is discontinuous at the frontiers of physical space, a feature that allows the transfer of the modification of the fields inside the TO medium to the wave that exits it. Because the transformed boundary is a combination of rotation and translations of the original boundary, the discontinuity does not cause reflections at the limits of the TO medium \cite{Schmiele}. Following the TO equations \eref{eq:parametersG} the EM parameters of the transformation medium are obtained:
\begin{equation}
\label{eq:BSparameters1}
\hat{\epsilon}_{TO}/\epsilon_0=\hat{\mu}_{TO}/\mu_0=
\left( \begin{array}{ccc}
    1&a&0\\a&1+a^2&0\\0&0&1
\end{array}\right)
\end{equation}
A 3D SPP parallel shifter should, in principle, be composed of a slab in the dielectric side sized $d\times l\times h_d$ with permitivitty $\hat{\epsilon}_{TO}$ and permeability $\hat{\mu}_{TO}$ along with a slab in the metal film of dimensions $d\times l\times h_m$ and parameters $\hat{\epsilon}=\hat{\epsilon}_{TO}\cdot \epsilon_m$ and $\hat{\mu}=\hat{\mu}_{TO}$. Again, the height of the slabs at each side must be larger than the SPP decay length in that medium. In this case, the TO medium parameters are anisotropic but homogeneous and take non-singular values. 

\Fref{fig:Shifter}(a) shows the transmittance spectrum resulting from 3D simulations. An SPP with a gaussian profile of width $\Delta=l-2ad$ propagates in the x direction and impinges on the transformation medium. Thus, it experiences a shift of $\phi=45^\circ$ (we set $a=1$) perpendicular to its propagation, in the $y$ direction, and exits the shifter propagating along a direction parallel to the original incident path. The in-plane size of the device is given by $l=20/3\lambda$ and $d=l/4$. The blue and red spectra in \fref{fig:Shifter}(a) show the transmittance through the beam shifter with and without introducing a transformation medium in the metal side, respectively. In the ideal situation (blue line), $100\%$ of the SPP powerflow is shifted over the whole spectra, whereas in the approximation (red line) the transmission is lowered due to the fact that the fraction of the SPP energy that is carried inside the metal is not shifted. At $600$ nm the transmission is lowered to $95.5\%$ while at $800$ nm amounts to $98.5\%$. On the other hand, for wavelengths larger than $900$ nm the red and blue lines differ by less than $1\%$ because most of the SPP energy is carried outside the metal. Therefore, manipulation of the metal is not required since the performance of the SPP shifter is very good when the TO parameters are introduced in the dielectric side in the whole optical range. The effect of the shifter on an SPP can be seen in \fref{fig:Shifter}(c), where the field distribution of an SPP at $\lambda=900$ nm undergoing a $45^\circ$ shift is shown. As a consequence of the shift, the outgoing propagation direction is $1.5$ $\mu$m from the incident path. Note that the SPP wave fronts nicely follow the vertical lines of the transformation map shown in (b).

\subsection{SPP right-angle bend}
\begin{figure}[h] \centering
\includegraphics{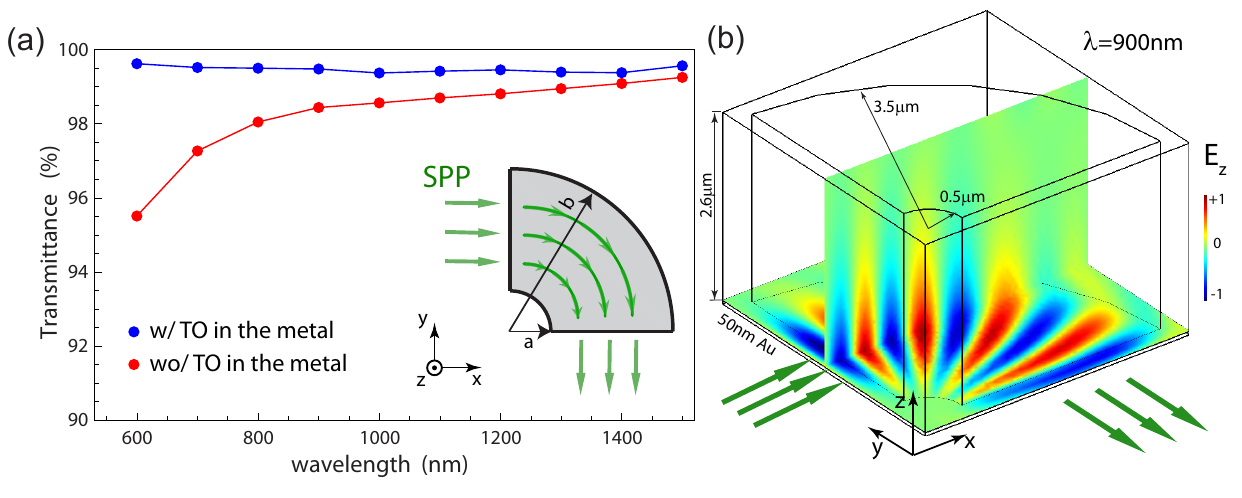} 
	\caption{3D simulation results of a SPP right-angle bend. (a) Transmission through the bend for SPPs of gaussian profiles (width $\Delta$) at different wavelengths. The blue and red lines show the case with and without placing a transformation medium in the metal side, respectively. The bend curvature radius is $\rho=5/2\lambda$ and the width is  $\Delta=10/3\lambda$. The inset panel shows the geometry layout from a top view. (b) z-component of the electric field of an SPP at $900$ nm travelling through the bend without applying TO in the metal side.}
    \label{fig:Bend}
\end{figure}

The next plasmonic element we consider here is a right-angle bend, a device that rotates the propagation direction of an incident SPP by $\frac{\pi}{2}$ (see the inset of \fref{fig:Bend}(a)). We start by describing a 2D beam bend, which is designed by transforming a a squared region of side $b$ in a Cartesian grid into a polar grid of squared cross-section $b\times b$ \cite{Kwon}. In order to bend a beam that propagates in the $x$ direction by $\frac{\pi}{2}$, so that it exits the device propagating in the $y$ direction, the appropriate transformation is the following:
\begin{equation} \label{eq:bend}
 r=y'\ \mbox{;} \ \ \phi=\frac{\pi}{2b}(b-x')\ \mbox{;} \ \  \ z=z'
\end{equation}
with $r=(x^2+y^2)^{1/2}$ and $\phi=\arctan(y/x)$. Again, this transformation is not continuous at the boundaries of the domain.The EM parameters retrieved from TO equations \eref{eq:parametersG} are expressed in Cartesian coordinates according to \eref{eq:CylCparameters} with:
\begin{equation}  \label{eq:Bendparameters}
 \epsilon_r=\frac{2b}{\pi r} \ \mbox{;}\ \ \epsilon_\theta=\frac{\pi r}{2b}\ \mbox{;}\ \ \epsilon_z=\frac{2b}{\pi r}
\end{equation} These tensors are singular at $r=0$ ( $\epsilon_{r}\mbox{,}\ \epsilon_{z} \to +\infty$ and $\epsilon_{\theta}\to0$). In order to avoid the divergence at $r=0$ we limit the extreme values of the parameters by limiting the radius to vary from an inner value $r=a$ to the outer value $r=b$. Then the width of the bend is $\Delta=b-a$ and the curvature radius is $\rho=(a+b)/2$. According to our methodology, to design an SPP right-angle bend we need to perform the coordinate transformation in the cylindrical shell comprised between $r=a$, $r=b$, $\phi=0$ and $\phi=\frac{\pi}{2}$. Ideally, we would need to implement the parameters $\hat{\epsilon}_{TO}$ and $\hat{\mu}_{TO}$ given by \eref{eq:CylCparameters} and \eref{eq:Bendparameters} within the dielectric material up to a height of $h_d$, while the metal optical properties should be modified following the expressions $\hat{\epsilon}=\hat{\epsilon}_{TO}\cdot \epsilon_m$ and $\hat{\mu}=\hat{\mu}_{TO}$ in a region of height $h_m$. 

\Fref{fig:Bend}(a) shows the transmittance curve of an SPP bend for two different situations: with and without implementing the TO parameters in the metal side (blue and red lines, respectively). The transmittance through the bend is calculated from simulations where SPPs at wavelengths ranging from $600$ to $1500$ nm are launched into the entrance of a bend of curvature radius $\rho=\frac{5}{2}\lambda$. The field distribution results from a simulation of an SPP at $\lambda=900$ nm propagating in the $x$ direction and travelling through a right angle bend of radius $\rho=1.5$ $\mu$m is shown in \fref{fig:Bend}(b). The transmittance is obtained from measurements of the power flow at the entrance and exit faces of the bend. The blue line in the transmittance curve corresponds to the ideal case where the TO parameters are implemented both in the metal and dielectric sides. The transmission is over $99\%$ for all wavelengths. On the other hand, the red line follows the transmittance through the bend when only the dielectric side is modified. It is clearly seen that this is indeed a very good approximation, as the transmission is always higher than $95\%$, approaching the same results as the ideal simulations at large wavelengths. For wavelengths larger than $900$ nm, the blue and red curves differ by only $2\%$ or less, meaning that by not structuring the metal side less than $2\%$ of the SPP powerflow will be lost. On the other hand, for short wavelengths ($600$-$800$ nm), the transmittance of the approximate situation is around $4\%$ lower than the transmittance of the ideal blue curve, a value that corresponds to the fraction of the SPP energy that is contained in the metal. Therefore, we have shown that in the approximate but more feasible situation of structuring only the dielectric side, the performance of the SPP bend is very good over the entire wavelength range considered. However, in some cases the SPP wave fronts might suffer some distortions even though the transmittance is still very high.

\subsection{SPP lens} \label{sect:Lens}
\begin{figure}[h] \centering
\includegraphics{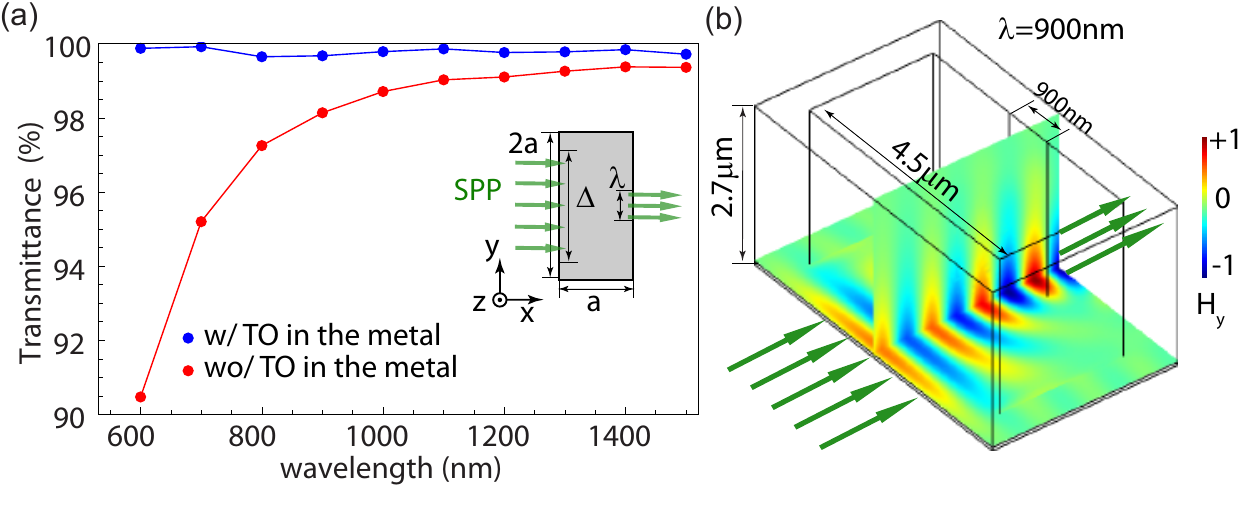} 
	\caption{Simulation results of a 3D SPP lens. An SPP exhibiting a gaussian profile of width $\Delta$ and propagating in the x direction impinges on the lens, where it is focused to a width of the order of the wavelength. (a) Transmittance spectra through the lens for two cases: with and without placing a transformation medium in the metal side (blue and red lines respectively). The transmittance is measured as the ratio of the power flow incident on the left to the power flow that exits within a wavelength width. The in-plane size of the lens is $a\times 2a$, with $a=2.5\lambda$, while the width of the incident beam is $\Delta=3.75\lambda$ (see inset panel). (b) 3D view of the y-component of the magnetic field of an SPP at $900$ nm travelling through the lens. In this simulation, the transformation medium was placed in the dielectric side only. Note that the lines parallel at the exit face separated by $900$ nm constitute an auxiliary element to measure the transmittance and do not represent a physical boundary.}
    \label{fig:Lens}
\end{figure}

Finally, we consider a plasmonic lens capable of focusing an incident gaussian SPP of a finite width. As described in \cite{Kwon}, transforming a semi-circular region of radius $a$ into a rectangular region sized $a \times 2a$ leads to a 2D rectangular lens whose focal point is located at the center of its exit face (see an schematic layout in the inset panel of \fref{fig:Lens}(a)). Assuming an incident wave from the left, we place the lens in the $x-y$ plane in the region defined by $-a\leq x\leq 0$ and $-a\leq y\leq a$. For such a geometry, the appropriate transformation is found to be:
\begin{equation} \label{eq:Lens}
 x=\frac{ax}{\sqrt{a^2-y^2}} \ \mbox{;} \ \  y=y' \ \mbox{;} \ \  z=z'
\end{equation}
The TO equations \eref{eq:parametersG} yield the following EM tensors in cartesian coordinates:
\begin{equation}\label{eq:Lensparameters}
\hat{\epsilon}_{TO}/\epsilon_0=\hat{\mu}_{TO}/\mu_0=
\left( \begin{array}{ccc}
   \frac{x^2y^2a^{-1}(a^2-y^2)^{-1}+a}{\sqrt{a^2-y^2}}&\frac{xy}{a\sqrt{a^2-y^2}} &0\\
    \frac{xy}{a\sqrt{a^2-y^2}}&\frac{\sqrt{a^2-y^2}}{a}&0\\
    0&0&\frac{\sqrt{a^2-y^2}}{a}
    \end{array}\right)
\end{equation}
These parameters are inhomogeneous and anisotropic and take singular values at the boundaries of the device. As in the previous examples, the lens for SPPs consists of a rectangular slab sized $a\times 2a \times h_d$ and placed on top of the metal film, with parameters $\hat{\epsilon}_{TO}$ and $\hat{\mu}_{TO}$. In principle, a rectangular slab of dimensions $a\times 2a \times h_m$ and parameters $\hat{\epsilon}_{TO}\cdot \epsilon_m$ and $\hat{\mu}_{TO}$ should be implemented in the metal side.

\Fref{fig:Lens} shows the results obtained from 3D simulations of a gaussian SPP of width $\Delta$ travelling through the lens. As a result, the SPP is focused at the exit side of the device and continues to propagate in cylindrical wavefronts. Panel a shows the transmittance through the device as a function of the SPP wavelength for a quantitative analysis of the lens effectiveness in two situations: when the TO parameters are or are not implemented in the metal side (blue and red curves, respectively). In the ideal case where transformation media are placed in the metal and dielectric sides, the transmittance is close to $100\%$ at all wavelengths. On the other hand, the red line shows that the approximation consisting of not modifying the metal side is very good for wavelengths above $900$ nm as in that wavelength range the transmittance lies above $98\%$. For wavelengths below $900$ nm the transmittance is lowered up to an $8\%$ compared to the ideal case because for shorter wavelengths there is a higher fraction of the SPP energy contained in the metal. However, the field patterns at these wavelengths (not shown here) reveal that the wave propagation and focusing is still very good. Panel b shows a plot of the field for an SPP at $900$ nm travelling through the lens. While the SPP travels a distance of $a=2250$ nm, its wavefronts are smoothly focused on a spot of the order of its wavelength.



\section{Conformal and quasiconformal mapping for realistic models} \label{sect:conformalAndQuasiconf}

The coordinate transformations we have considered so far present some technical problems when it comes to practical realizations as they generally yield highly anisotropic $\hat{\epsilon}$ and $\hat{\mu}$. However, if the we choose an appropriate coordinate transformation by imposing some mathematical requirements, isotropic EM parameters may be obtained. This means that only isotropic dielectrics would be necessary to construct devices, reducing losses and opening up the possibility of broad-band performances. In particular, conformal and quasiconformal transformations give rise to isotropic material parameters, yielding easier implementations of transformation media based on isotropic dielectrics. Conformal mappings were first investigated in \cite{Leonhardt06}, where an invisibility device was designed by applying a transformation in the complex plane to Helmholtz's equation. This concept was also applied to design other isotropic devices, such as arbitrary waveguide bends within the limit of geometrical optics \cite{Landy,Mei09} or directional antennas, flat lenses and beam bends \cite{Schmiele}. Regarding plasmonic systems, conformal and quasiconfomal transformations have been employed to design of light-harvesting \cite{Aubry} devices and ground-plane cloaks \cite{Liu10}.  
   
In the following we will show how to derive the isotropic EM parameters that correspond to a conformal transformation within the TO framework. Let us start by assuming that the virtual and physical domains are contained in the complex planes  $z=x'+iy'$ and $w=x+iy$, respectively. As we are dealing with SPPs, 2D waves extending into the third dimension (the direction perpendicular to the metal-dielectric interface), we can restrict our study to 2D transformations. A 2D coordinate transformation from virtual to physical coordinates can then be described by means of a complex function:
\begin{equation}
	w=f(z,\overline{z})=x(x',y')+i y(x',y')
\end{equation} 
This transformation is said to be conformal if the function $f(z)$ depends on $z$ but not on its conjugate $\overline{z}$ and is analytical, i.e. $f'(z)$ can be defined. Then, the Cauchy-Riemann equations are satisfied ($ \partial x/\partial x'=\partial y/\partial y'$ and $\partial x/\partial y'=-\partial y/\partial x'$). Consequently, the metric tensor of a conformal transformation reduces to a diagonal matrix:
\begin{equation}
\widehat{G}^{-1}=g^{11}\cdot\mathbb{I}_{2\times2} \ \ \ \mbox{where} \ \ 	g^{11}=(g_{11})^{-1}=\frac{\partial^2 x}{\partial x'^2}+\frac{\partial^2 x}{\partial y'^2}
\end{equation}
This means that the grid described by the transformation map $w=f(z)$ is orthogonal, i.e., the angles between coordinate lines are preserved after the mapping. According to the TO rule \eref{eq:parametersG}, we calculate the EM parameters of the transformation medium that corresponds to a 2D conformal transformation:
\begin{equation} \label{eq:conformalparam}
	\hat{\epsilon}_{TO}/\epsilon_0=\hat{\mu}_{TO}/\mu_0=\sqrt{g} \widehat{G}^{-1}=\left(\begin{array}{ccc} 1 & 0 & 0 \\ 0 &1 & 0 \\ 0 & 0 & \sqrt{g}
 \end{array} \right) 
\end{equation}
Now let us consider the case of an EM wave with the electric field polarized along the z-axis and the magnetic field in the x-y plane. The propagation of this wave is determined by the electric permittivity along the z-axis and the magnetic susceptibility tensor in the x-y plane. According to the equation above, these are the following:
\begin{equation} 
		\epsilon_z=\epsilon_0\sqrt{g} \ \ \mbox{;} \ \ \ \hat{\mu}=\mu_0 \mathbb{I}_{2\times2}
\end{equation}
This set of parameters is equivalent to an isotropic refractive index:
\begin{equation} \label{eq:n_isotropic}
	n=\sqrt{\epsilon_z\mu}=n_0 g^{1/4}
\end{equation}
where $n_0=\sqrt{\epsilon_0\mu_0}$ is the background refractive index. If on the other hand, the EM wave is polarized with the magnetic field along the z-axis, then $\epsilon=\epsilon_0$, $\mu_z=\mu_0\sqrt{g}$ and $n=\sqrt{\epsilon \mu_z}=n_0g^{1/4}$. Notice that, as shown in the Appendix, our approach leads to the same result as the \emph{Optical Conformal Mapping} technique based on Helmholtz's equation \cite{Leonhardt06}. Therefore, for problems where the two polarizations can be treated separately, a 2D conformal transformation leads to an isotropic transformation medium, with a refractive index defined by \eref{eq:n_isotropic}. On the other hand, SPPs need to be considered separately as their dispersion relation depends on the electric permittivity and the magnetic permeability rather than on the refractive index. However, we will show in the following that the propagation of an SPP in a medium with the exact diagonal anisotropic parameters \eref{eq:conformalparam} is very similar to that in a medium with the approximate isotropic refractive index \eref{eq:n_isotropic} for a given conformal transformation. 

Thus, conformal transformations lead to transformation media described by isotropic refractive indexes. However, an analytical conformal transformation is not known for every desired functionality. In those cases, quasiconformal maps are useful for designing TO media based on isotropic materials. They constitute a subset of coordinate transformations, $w=f(z,\overline{z})$, that satisfy less requirements than conformal transformations. In particular, they can depend on $z$ and $\overline{z}$. In contrast to conformal transformations, quasiconformal transformations map squares in the original space to rectangles of constant aspect ratio in the transformed space. Since they are generated numerically through the minimization of, for instance, the {\itshape Modified-Liao} or {\itshape Winslow} functionals \cite{Knupp94,GridGeneration}, they are able to treat arbitrarily complex boundaries. A quasiconformal map is the optimal map of minimum anisotropy for a given domain, a fact that was can be used to design isotropic transformation media. In reference \cite{JensenLi08}, a quasiconformal map for a ground-plane cloak was obtained by minimizing the {\itshape Modified-Liao} functional with sliding boundary conditions. Applying this idea, several invisibility ground-plane cloaks operating at optical \cite{Valentine09,Gabrielli} and microwave \cite{Liu09} frequencies were fabricated out of isotropic dielectric materials and a non-resonant metallic metamaterial, respectively. 

In this work, we have made use of a recently proposed \cite{Chang,Hu} approach to quasiconformal mapping generation that does not need of functional minimizations. This method takes advantage of the fact that the transformation functions, $x'(x,y)$ and $y'(x,y)$, that minimize {\itshape Winslow} functional, also solve Laplace's equation:
\begin{equation} \label{eq:LaplaceInv}
	\nabla^2 x'=0  \mbox{;} \ \ \ \nabla^2 y'=0
\end{equation}
where the Laplacian, taken in the physical space, is acting on the virtual coordinates. Therefore, solving Laplace's equation in physical domain subject to sliding boundary conditions, $\vec{n}\cdot \vec{\nabla}x'^i$, leads to a quasiconformal map. Once the functions $x'(x,y)$ and $y'(x,y)$ are obtained, the gradients of the transformation can be computed in order to construct the inverse Jacobian matrix, $\widehat{A}^{-1}=\partial x^{i'}/\partial x^i$. By inverting it, the Jacobian matrix needed for the TO procedure is obtained. Now, assuming that the anisotropy of the quasiconformal mapping is negligible, we can approximate its metric tensor by a diagonal matrix and define the transformation medium by means of \eref{eq:conformalparam}. Again, for problems where the polarization state of the EM wave is well defined an isotropic refractive index can be defined following \eref{eq:n_isotropic}. In this way, a quasiconformal transformation can be used to design an isotropic TO medium defined by the index profile $n=g^{1/4}$.

\subsection{Isotropic implementation of an SPP shifter}
\begin{figure}[h] \centering
\includegraphics{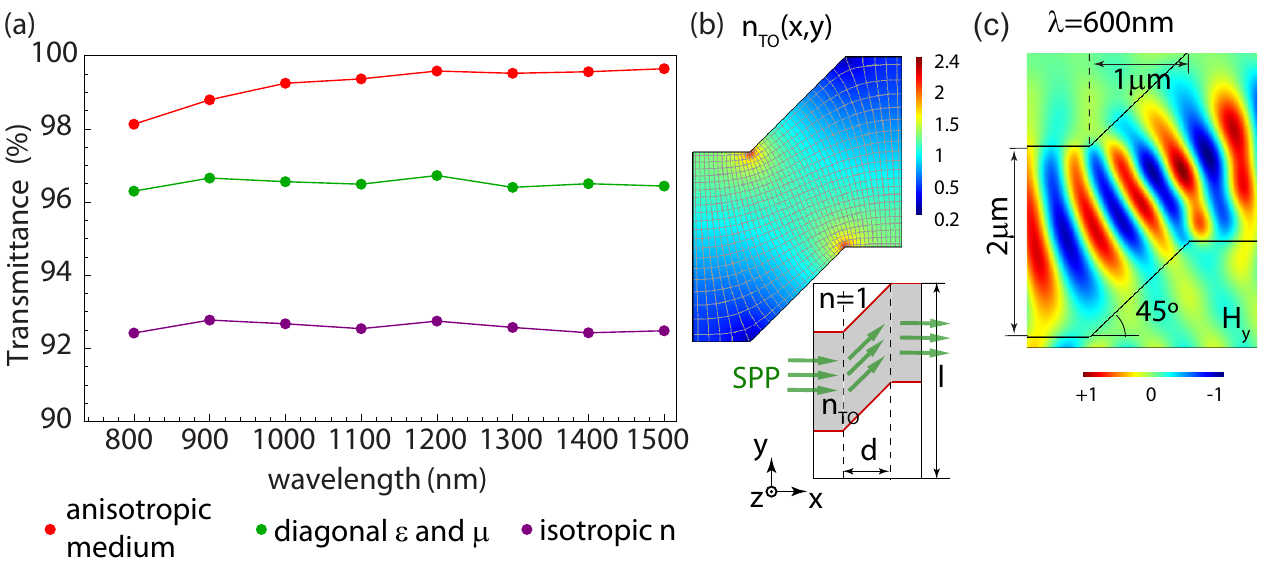} 
	\caption{Simulation results of an isotropic SPP parallel shifter. (a) Transmittance through the shift for SPPs at different wavelengths. The red line shows the transmittance through a transformation medium characterized by anisotropic parameters (same as in \fref{fig:Shifter}). The green and purple lines correspond to a parallel shifter characterized by the diagonal $\hat{\epsilon}_{TO}$ and $\hat{\mu}_{TO}$ and the isotropic refractive index, respectively, both retrieved from the quasiconformal transformation. For all the cases, the TO parameters are implemented in the dielectric side only. (b) Quasiconformal map (grid) and refractive index profile (color map) of the device. The inset panel show the geometry layout. (c) Magnetic field, $H_y$, of an SPP at $600$ nm travelling through the isotropic shifter after implementing the index shown in panel b.}
    \label{fig:ShifterIsotropic}
\end{figure}

As a first example of a plasmonic device based on isotropic dielectrics, let us consider the design of an SPP parallel shifter using a quasiconformal mapping. As in the previous examples, a TO device will be placed on top of the surface through which the SPP travels. In order to design an isotropic shifter with the same functionality as the anisotropic one previously considered, we start by defining the geometry shown in the inset panel of \fref{fig:ShifterIsotropic}(b). 
The transformation medium, whose top view is shown in gray, is divided into three regions: an entrance for the SPP, of width $l/2$, at the left, the shifting region (at $45^\circ$), of length $d$, and an exit region at the right. Then, we solve Laplace's equations within the gray region with sliding boundary conditions $\vec{n}\cdot \vec{\nabla}x'$ at the lower and upper boundaries and $\vec{n}\cdot \vec{\nabla}y'$ at the left and right boundaries. The resulting quasiconformal map is plotted in \fref{fig:ShifterIsotropic}(b), where the grid corresponds to lines of $x'(x,y)=constant$ and $y'(x,y)= constant$. Following the method described above, we obtain an isotropic refractive index profile that varies between $0$ and $3.5$ (color map in \fref{fig:ShifterIsotropic}(b)). 

The transmittance curve in \fref{fig:ShifterIsotropic}(a) shows the results obtained from 3D simulations of the isotropic SPP shifter. SPPs at different wavelengths (from $800$ to $1500$ nm) travel along a gold-vacuum interface and impinge on the transformation medium placed on top of the gold surface. As we have previously shown, a transformation medium in the metal is not needed. The size of the device is the same as for the anisotropic shifter discussed above. Three different transmittance spectra have been calculated. The green and purple lines show the transmittance through the transformation medium resulting from the quasiconformal transformation. While the green line corresponds to the diagonal $\hat{\epsilon}_{TO}$ and $\hat{\mu}_{TO}$ given by \eref{eq:conformalparam}, the purple line corresponds to the index profile $n(x,y)$ calculated according to \eref{eq:n_isotropic} and shown in panel b. On the other hand, the red line shows the transmittance of the anisotropic shifter (same as in \fref{fig:Shifter}), so that the efficiency of both approaches can be compared. Whereas the red line takes transmission values above $98\%$, the green and purple lines lie beneath, around $96\%$ and $92\%$ respectively. The are two main reasons for this feature. First, the fact that the transformation is quasiconformal leads to an impedance mismatch at the interfaces to free space, causing losses and lowering the green and purple curves. Second, the purple curve shows the lowest values in transmittance because when $\hat{\epsilon}_{TO}$ and $\hat{\mu}_{TO}$ are replaced by an isotropic index, the dispersion relation of the SPP is affected. In this case, the wave fronts of the SPP are distorted in order to follow the vertical lines of the quasiconformal map shown in panel b. This feature can be seen in \fref{fig:ShifterIsotropic}(c), were an SPP at $600$ nm is guided through an isotropic beam shifter. Despite the fact that an isotropic index reduces the transmittance, the shifting of the SPP is nicely accomplished.

\subsection{Isotropic implementation of an SPP right-angle bend}
\begin{figure}[h] \centering
\includegraphics{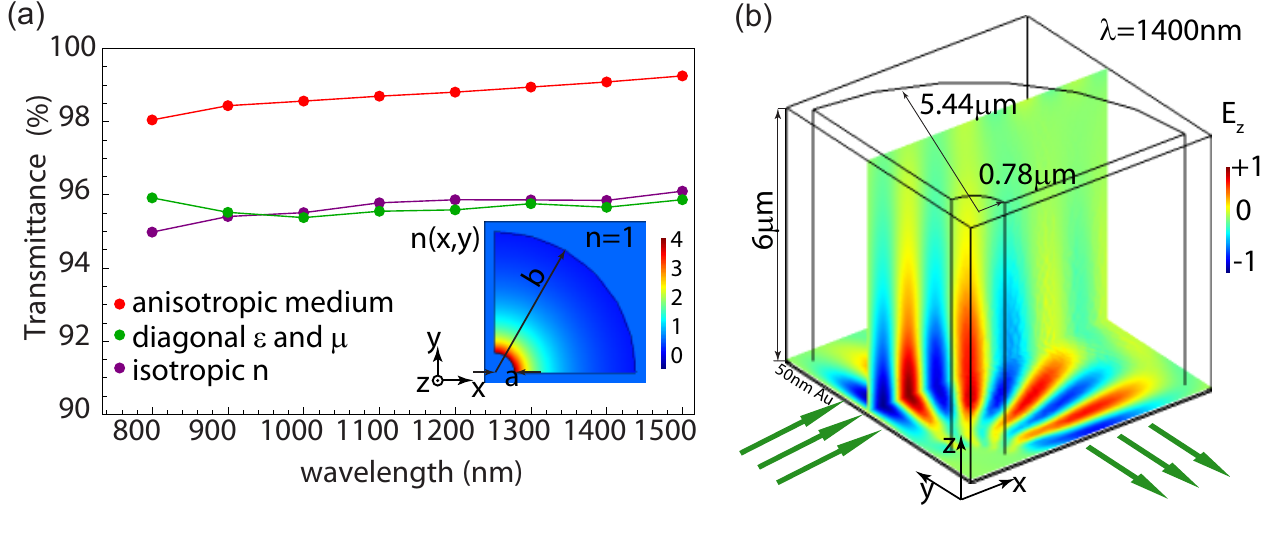} 
	\caption{Simulation results of an SPP bend characterized by an isotropic refractive index. (a) Transmittance through the bend for SPPs at different wavelengths comparing bends with anisotropic, diagonal and isotropic parameters. The red line corresponds to the case of a transformation medium characterized by anisotropic parameters (same as in \fref{fig:Bend}). The green and purple lines stand for a conformal transformation: green curve for transmittance resulting from diagonal parameters and purple curve for the isotropic refractive index (top view plotted in the inset panel). In all cases, the TO parameters are implemented in the dielectric side only. The geometrical parameters are the same as in \fref{fig:Bend}. (b) Electric field, $E_z$, of an SPP at $1400$ nm travelling through the isotropic bend.}
    \label{fig:BendIsotropic}
\end{figure}

A beam bend of the same characteristics as that discussed in section 2.3 can also be realized with isotropic materials. By means of a conformal transformation, a more realistic optical device that bends the SPP propagation direction by a right-angle can be designed. The anisotropic parameters \eref{eq:Bendparameters}, yielded by the transformation \eref{eq:bend}, are replaced by an isotropic refractive index if a conformal transformation is made \cite{Schmiele}:
\begin{equation}  \label{eq:Bendindex}
 n(r)=\frac{2b}{\pi r}
\end{equation}
As illustrated in the inset of \fref{fig:BendIsotropic}(a), the transformation medium, characterized by $n(r)$, is a cylindrical shell, $a<r<b$, of height $h_d$ placed on top of the metal surface. This refractive index smoothly decreases from $4.46$ at $r=a$ to $0.67$ at $r=b$. 

We have tested the functionality of the isotropic SPP bend by means of 3D simulations. The curves in \fref{fig:BendIsotropic}(a) compare the transmittance through the SPP bend with anisotropic parameters (red line) and bends characterized by parameters resulting from a conformal transformation (green and purple lines) over a wavelength range between $800$ and $1500$ nm. The green line shows the transmittance through a TO medium characterized by the diagonal $\hat{\epsilon}_{TO}$ and $\hat{\mu}_{TO}$ calculated from the conformal transformation according to \eref{eq:conformalparam}. The $\epsilon_{zz}$ and $\mu_{zz}$ components of the tensors vary throughout the medium from $0$ to $16$. The purple line corresponds to a bend characterized by the isotropic refractive index given by \eref{eq:Bendindex}. In the last two situations the transmittance is limited to $96\%$ due to impedance mismatch to free space. This feature, not present for the anisotropic case (red curve), arises when conformal transformations are used. As shown in the inset panel of \fref{fig:BendIsotropic}(a), the refractive index abruptly changes from $n=1$ to $n(r)$ at the interfaces along the entrance and exit faces of the bend. As a consequence, the SPP suffers reflections that lower the transmission. In this case, the green and purple lines lie very close to each other, meaning that the dispersion relation of the SPP is not affected by changing $\epsilon_{zz}$ and $\mu_{zz}$ to $n$. The main reason for this is that the parameters resulting from the conformal transformation are smoother and lead to less distorted wavefronts than the quasiconformal transformation in the previous example of an SPP shifter does. \Fref{fig:BendIsotropic}(b) illustrates the performance of the bend for a gaussian SPP at $1400$ nm. After impinging on the device, close to $96\%$ of the SPP is bent in a right angle and leaves the transformation medium through the exit face to continue its propagation through the metal.

\subsection{Isotropic implementation of an SPP lens}
\begin{figure}[h] \centering
\includegraphics{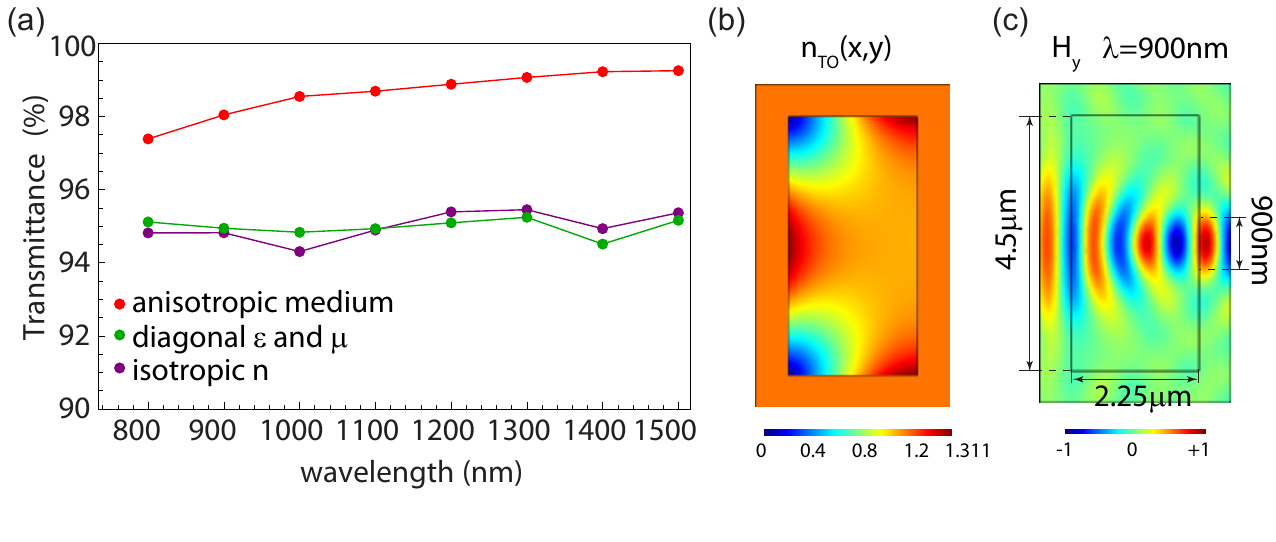} 
	\caption{Results obtained from 3D simulations on SPPs propagating through a lens. (a) Transmittance spectra through lenses characterized by different EM parameters. Red line: anisotropic $\hat{\epsilon}_{TO}$ and $\hat{\mu}_{TO}$ (same as in \fref{fig:Lens}). Green line: diagonal $\hat{\epsilon}_{TO}$ and $\hat{\mu}_{TO}$ resulting from a conformal transformation according to \eref{eq:conformalparam}. Purple line: isotropic refractive index resulting from the same conformal transformation. (b) Refractive index profile yield by the conformal transformation. (c) Top-view of the y-component of the magnetic field. An SPP travels through a lens characterized by the refractive index shown in b and is focused within a width of the order of the wavelength. }
    \label{fig:LensIsotropic}
\end{figure}

In this section we discuss the realization of an SPP lens with isotropic materials. The geometry and purpose of the device are the same as presented previously (see section 2.4). An isotropic refractive index is obtained if, 
instead of the transformation given by \eref{eq:Lens}, a conformal transformation between a semicircle and a rectangle is performed. In particular, we will make use of a Schwarz-Christoffel transformation as described in \cite{Schmiele}. The refractive index profile that results from this transformation smoothly varies from $0$ to $1.31$ and is shown in \fref{fig:LensIsotropic}(b). To design the SPP lens a transformation medium with a rectangular cross-section characterized by the index distribution $n_{TO}(x,y)$ and height $h_d$ is placed on top of a metal film. The device is surrounded by vacuum ($n=1$).

\Fref{fig:LensIsotropic} shows the results obtained from 3D simulations of SPPs propagating through the lens. The three transmittance curves in panel a compare the functionality of SPP lenses characterized by different EM parameters. The red line shows the best performance as it represents the transmittance through a lens characterized by the anisotropic and singular parameters given by \eref{eq:Lensparameters} (see \fref{fig:Lens}). The green and purple lines correspond to lenses designed by means of a conformal transformation and therefore are situated at lower transmittance values due to impedance mismatch to free space. The green line accounts for the transmittance through a lens with diagonal parameters, $\hat{\epsilon}_{TO}$ and $\hat{\mu}_{TO}$, that are obtained from the conformal transformation according to \eref{eq:conformalparam}. Only the third component of the tensors diagonal is different from $1$, and it varies smoothly from $0$ to $1.72$. The purple line corresponds to the isotropic refractive index retrieved from the conformal transformation according to \eref{eq:n_isotropic} and is shown in panel b. The two cases considered for the conformal transformation (diagonal $\hat{\epsilon}_{TO}$ and $\hat{\mu}_{TO}$ and isotropic $n$) give close transmittance results in this example, similarly to the SPP bend results above. Panel c shows the field pattern of an SPP travelling trough the isotropic lens as its wavefronts are focused on the exit face.     

\subsection{SPP ground-plane cloak}
\begin{figure}[ht] \centering
\includegraphics{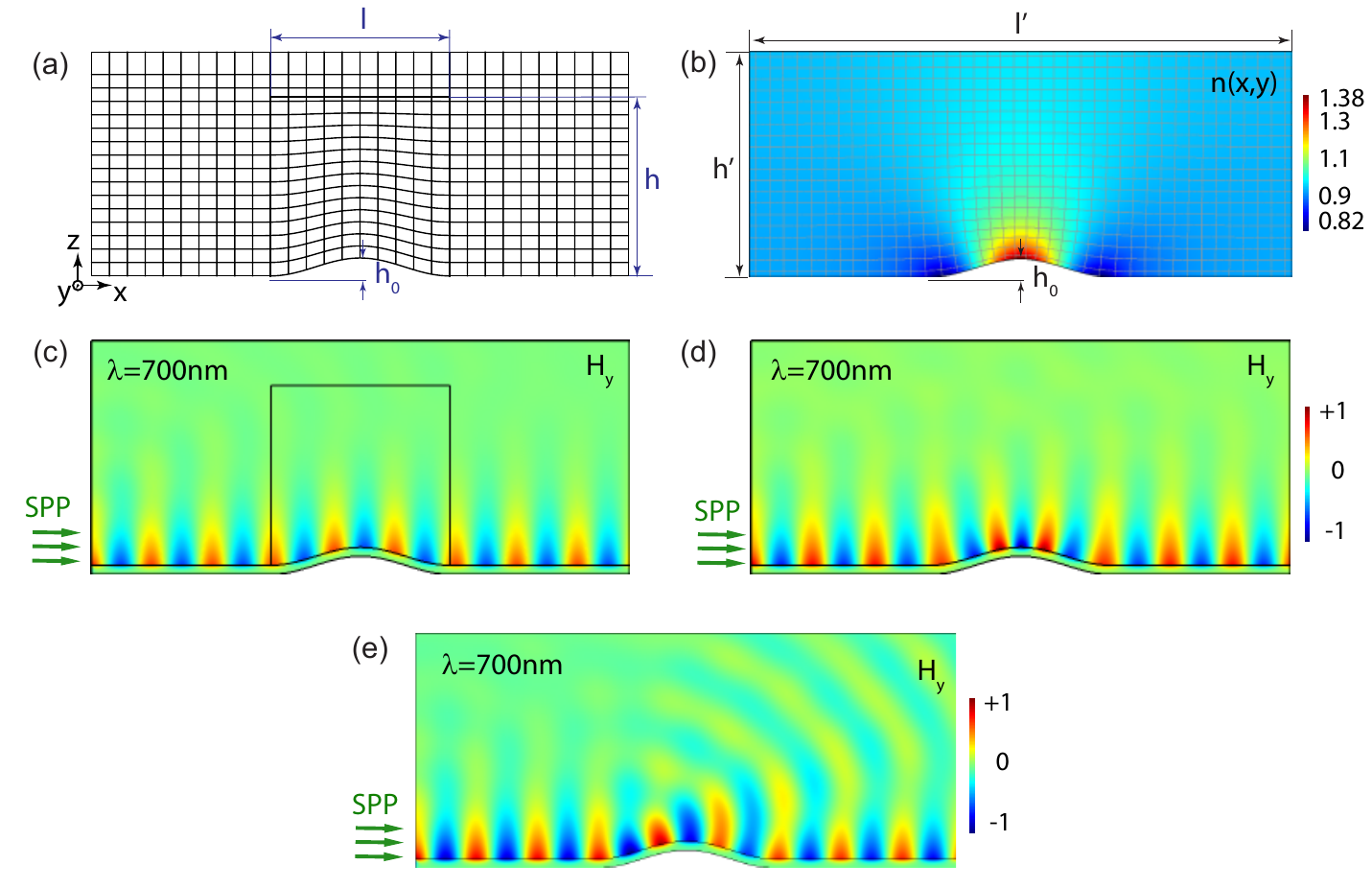} 
	\caption{Ground-plane cloak for a bump of height $h_0=0.2$ $\mu$m on a gold-vacuum interface infinite in the $y$ direction. (a) Coordinate map of the transfinite transformation that leads to a ground-plane cloak characterized by anisotropic $\hat{\epsilon}_{TO}$ and $\hat{\mu}_{TO}$. The cloak, of length $l=2$ $\mu$m and height $h=2$ $\mu$m, is placed in the dielectric side of the interface. (b) Refractive index profile (color map) resulting from the quasiconformal map shown by the grid. The size of cloak is given by $l'=6$ $\mu$m and $h'=2.5$ $\mu$m. (c) and (d) Magnetic field, $H_y$, of an SPP at $700$ nm travelling through the anisotropic (a) and the isotropic (b) cloaks, respectively. (e) SPP scattering on a gold bump when the cloak is not present.}
    \label{fig:GroundPlaneCloak}
\end{figure}

Finally, let us compare the two approaches to TO that we have presented (anisotropic and isotropic transformation media) as they apply to a ground-plane cloak for SPPs. This cloak was first proposed for mimicking a flat reflective surface under plane wave illumination \cite{JensenLi08}. Given that TO is valid for any kind of EM waves, a ground-plane cloak can also be used to suppress the scattering of an SPP from a bump on the metallic surface along which the plasmon propagates (see \fref{fig:GroundPlaneCloak}). Thus, the SPP appears to have travelled along a flat metal surface although it is physically travelling along the bump. Let us consider an SPP propagating in the $x$ direction along a surface placed in the $x$-$y$ plane. In order to cloak a bump defined by the equation $z(x)=h_0\cos^2\left(\frac{\pi}{l}x\right)$, we make a transformation in the $x$-$z$ plane (perpendicular to the interface) from a rectangular region sitting on a metal surface to a region whose bottom boundary is lifted following the surface of the bump. We have considered two different transformations that yield the same functionality: a transfinite transformation, \fref{fig:GroundPlaneCloak}(a), and a quasiconformal mapping, \fref{fig:GroundPlaneCloak}(b). 

The transfinite transformation shown in panel a is made in the dielectric side of the interface in a box sized $l \times h$ and has an analytical expression (details can be found in \cite{Huidobro}). Since it is not a conformal transformation, the EM parameters, $\hat{\epsilon}_{TO}$ and $\hat{\mu}_{TO}$, retrieved for the corresponding transformation medium are anisotropic. When the cloak is placed on top of the surface, an SPP smoothly travels across the bump without any scattering. \Fref{fig:GroundPlaneCloak}(c) renders the magnetic field component, $H_y$, of an SPP at $700$ nm travelling through a bump of height $0.2$ $\mu$m. On the other hand, we have calculated the isotropic index profile resulting from the quasiconformal map shown in panel b. In this case, the transformation medium is sized $h' \times l'$. The quasiconformal map was generated by solving Laplace's equation with sliding boundary conditions $\vec{n}\cdot \vec{\nabla}x'$ (at the lower and upper boundaries) and $\vec{n}\cdot \vec{\nabla}y'$ (at the left and right boundaries). The effect of a ground-plane cloak is achieved by setting $y'=0$ at the lower boundary so that the bump is mapped onto a flat surface. Then we have calculated the refractive index following the procedure described above. It ranges from $0.82$ to $1.38$ and the background value ($n=1$) is recovered at the outer boundaries of the cloak. This means that the cloak is impedance-matched to vacuum, a fact that suppresses possible reflections at the interfaces. An SPP travelling through the isotropic cloak is illustrated in panel d, where the SPP smoothly passes the bump. In both cases (panels c and d), the wave fronts follow the transformed grid. In contrast, if no cloak is placed on top of the bump, the SPP suffers strong scattering as shown in \fref{fig:GroundPlaneCloak}(e).

\section{Conclusions} \label{sect:conclusions}
In conclusion, we have demonstrated that the application of the concept of Transformation Optics to Plasmonics brings into the field unprecedented control over the flow of SPPs. Based on this approach, we have presented the necessary tools to design feasible plasmonic devices. When considering SPPs propagating on metal-dielectric interfaces, a direct application of Transformation Optics would imply a manipulation of the optical properties of both the metal and dielectric materials. However, we have shown that modifying only the dielectric side leads to almost perfect functionalities because most of the SPP energy is carried within the dielectric medium at optical frequencies. Having proved this feature within a general framework in which the EM parameters are anisotropic, we have then presented plasmonic devices that can be made out of isotropic dielectric materials. Making use of conformal and quasiconformal transformations, control over SPPs flow can be achieved by placing a structured dielectric layer on top of the metal surface. Regarding practical implementations, this fact would render low loss and broadband performances. For instance, holes could be drilled in a dielectric layer to effectively mimic an isotropic refractive index. We expect that the flexibility of our methodology will ensure many different theoretical proposals and experimental realizations of plasmonic devices based on Transformation Optics.

\ack This work has been sponsored by the Spanish Ministry of Science under projects MAT2009-06609-C02 and CSD2007-046-NanoLight. P. A. H. acknowledges financial support from the Spanish Ministry of Education through grant AP2008-00021. 

\newpage
\appendix
\section{Optical conformal mapping}\label{Ap}

In the limit of ray optics, where the refractive index $n(x,y)$ is nearly constant over the scale of a wavelength $\frac{2 \pi}{\left|\vec{k}\right|}$, the field amplitude for each polarization satisfies Helmholtz's equation.
In virtual space ($z=x'+iy'$), this equation reads:
\begin{equation}
	\left(\nabla'^2+\frac{\omega^2}{c^2}n'^2\right)\psi=0
\end{equation}
where the Laplacian is $\nabla'^2=\partial^2/\partial x'^2+\partial^2/\partial x'^2$.
Let us now consider a transformation $w=w(z)$ from virtual to physical space ($w=x+iy$) such that it is conformal. Making use of the relation between the Laplacian in virtual and physical spaces, $\nabla'^2=\left|d \omega/d z\right|^2 \nabla^2$, we can transform Helmholtz's equation to physical space:
\begin{equation}
	\left(\nabla^2+\frac{\omega^2}{c^2}\left|\frac{d z}{d \omega}\right|^2n'^2\right)\psi=0 
\end{equation}
This equation is left invariant if we choose a refractive index of the form \cite{Leonhardt06}:
\begin{equation}
n=\left|\frac{d z}{d \omega}\right|n'
\end{equation}
where $n'$ is the background refractive index, $n_0$. By means of the Cauchy-Riemman equations we can express the refractive index in terms of the metric tensor of a conformal transformation:
\begin{eqnarray} 
		n=n_0\left|\frac{d}{d \omega}(x'+iy')\right|=n_0\left|\frac{\partial x'}{\partial x}+i\frac{\partial y'}{\partial x} \right|=n_0\sqrt{\frac{\partial^2 x'}{\partial x^2}+\frac{\partial^2 y'}{\partial x^2}}=n_0g^{1/4}
\end{eqnarray}
which is the same expression for the refractive index obtained according to TO \eref{eq:n_isotropic}.

\end{document}